\documentclass[letterpaper, 10 pt, conference]{ieeeconf}
\IEEEoverridecommandlockouts
\usepackage{cite}
\usepackage{amsmath,amssymb,amsfonts}
\usepackage{algorithmic}
\usepackage{graphicx}
\usepackage{textcomp}
\usepackage{xcolor}
\usepackage{url}
\usepackage{siunitx}
\usepackage{mathrsfs}
\usepackage{bm}
\usepackage{mathtools}
\usepackage{siunitx}
\def\BibTeX{{\rm B\kern-.05em{\sc i\kern-.025em b}\kern-.08em
    T\kern-.1667em\lower.7ex\hbox{E}\kern-.125emX}}

\usepackage[acronym,shortcuts]{glossaries}
\newacronym{GPR}{GPR}{Gaussian process regression}
\newacronym{MAE}{MAE}{mean absolute error}
\newacronym{PDS}{PDS}{Poisson disk sampling}
\glsdisablehyper

\begin{document}

\title{\LARGE \bf Efficient Spatial Estimation of Perceptual Thresholds for Retinal Implants via Gaussian Process Regression

\author{Roksana Sadeghi$^{1}$ and Michael Beyeler$^{1,2}$}
\thanks{$^{1}$RS and MB are with the Department of Computer Science, University of California, Santa Barbara, CA 93106, USA.}
\thanks{$^{2}$MB is with the Department of Psychological \& Brain Sciences, University of California, Santa Barbara, CA 93106, USA.}
\thanks{The authors confirm contribution to the paper as follows. 
Study conception and design: MB; 
data collection: RS;
computational model development: MB; 
data analyis: RS, MB;
manuscript writing: RS, MB;
all approved the final version of the manuscript.}
\thanks{Supported by the National Library of Medicine of the National Institutes of Health (NIH) under Award Number DP2LM014268. The authors would like to thank Tori LeVier for her support in recruiting and managing participants. The content is solely the responsibility of the authors and does not necessarily represent the official views of the NIH. }
}

\maketitle

\begin{abstract}
Retinal prostheses restore vision by electrically stimulating surviving neurons, but calibrating perceptual thresholds (i.e., the minimum stimulus intensity required for perception) remains a time-intensive challenge, especially for high-electrode-count devices. Since neighboring electrodes exhibit spatial correlations, we propose a Gaussian Process Regression (GPR) framework to predict thresholds at unsampled locations while leveraging uncertainty estimates to guide adaptive sampling.
Using perceptual threshold data from four Argus II users, we show that GPR with a Matérn kernel provides more accurate threshold predictions than a Radial Basis Function (RBF) kernel (\bm{$p < .001$}, Wilcoxon signed-rank test). In addition, spatially optimized sampling yielded lower prediction error than uniform random sampling for Participants 1 and 3 (\bm{$p < .05$}). While adaptive sampling dynamically selects electrodes based on model uncertainty, its accuracy gains over spatial sampling were not statistically significant (\bm{$p > .05$}), though it approached significance for Participant 1 (\bm{$p = .074$}).
These findings establish GPR with spatial sampling as a scalable, efficient approach to retinal prosthesis calibration, minimizing patient burden while maintaining predictive accuracy. More broadly, this framework offers a generalizable solution for adaptive calibration in neuroprosthetic devices with spatially structured stimulation thresholds, paving the way for faster, more personalized system fitting in future high-channel-count implants.
\end{abstract}

\vspace{1ex}
\small
\textit{\textbf{Clinical Relevance---}}\textbf{Gaussian Progress Regression offers a scalable path toward faster, more personalized calibration procedures for future high-channel-count neuroprosthetic devices.}


\section{Introduction}

Retinal degenerative diseases, such as retinitis pigmentosa and age-related macular degeneration, lead to the progressive loss of photoreceptors, ultimately resulting in profound vision impairment. 
Visual prostheses offer a potential intervention by electrically stimulating the remaining retinal neurons to evoke visual percepts~\cite{weiland_electrical_2016}. 
However, achieving functional vision with these devices requires careful calibration of stimulation parameters, particularly the perceptual thresholds; that is, the minimum stimulus amplitude required to elicit a visual response.

Traditionally, perceptual thresholds are determined through extensive psychophysical testing, where stimuli of varying amplitudes are presented in a trial-by-trial manner, and patients report their perceptual experiences~\cite{de_balthasar_factors_2008,ahuja_factors_2013}. 
Although effective, this approach is time-consuming and labor-intensive, requiring tens of trials per electrode and frequent recalibrations due to threshold drift over time~\cite{yue_ten-year_2015,hu_explainable_2021}. 
In addition, thresholds exhibit substantial variability, both across subjects and among electrodes within the same implant, due to factors such as electrode-neuron distance, impedance, and local retinal health~\cite{de_balthasar_factors_2008,ahuja_factors_2013,shivdasani_factors_2014,pogoncheff_explainable_2024}. 
This variability makes calibration (``system fitting") an ongoing clinical challenge.
Recent work has also highlighted the importance of aligning research priorities with the real-world experiences and usability needs of prosthesis users~\cite{nadolskis_aligning_2024}, further motivating the development of calibration strategies that minimize patient burden, especially as future prostheses aim to scale from dozens to hundreds or thousands of electrodes~\cite{palanker_photovoltaic_2020,chenais_photovoltaic_2021,musk_integrated_2019}.

Current calibration strategies attempt to reduce the number of trials per electrode using adaptive psychophysical methods. 
For example, the Argus II epiretinal implant (60 electrodes)~\cite{luo_argusr_2016} employs a hybrid threshold estimation approach, combining the maximum likelihood method to adjust the stimulus range with the method of constant stimuli to estimate the perceptual threshold, requiring between 30 and 90 trials per electrode~\cite{sadeghi_comparing_2021}.
Bayesian adaptive procedures~\cite{sadeghi_using_2020,kontsevich_bayesian_1999} further reduce the number of trials to 11–30 per electrode, but this still results in over 600–1,800 trials per session, presenting a substantial burden to users.

Critically, perceptual thresholds are not independent across electrodes. Due to the spatial arrangement of retinal implants, neighboring electrodes often exhibit correlated thresholds, reflecting shared neural activation zones, current spread, and biological factors. This spatial structure suggests that a more principled, model-driven approach could infer unsampled thresholds by leveraging correlations, dramatically reducing the number of measurements needed.

\begin{figure*}[!th]
    \centering
    \includegraphics[width=\linewidth]{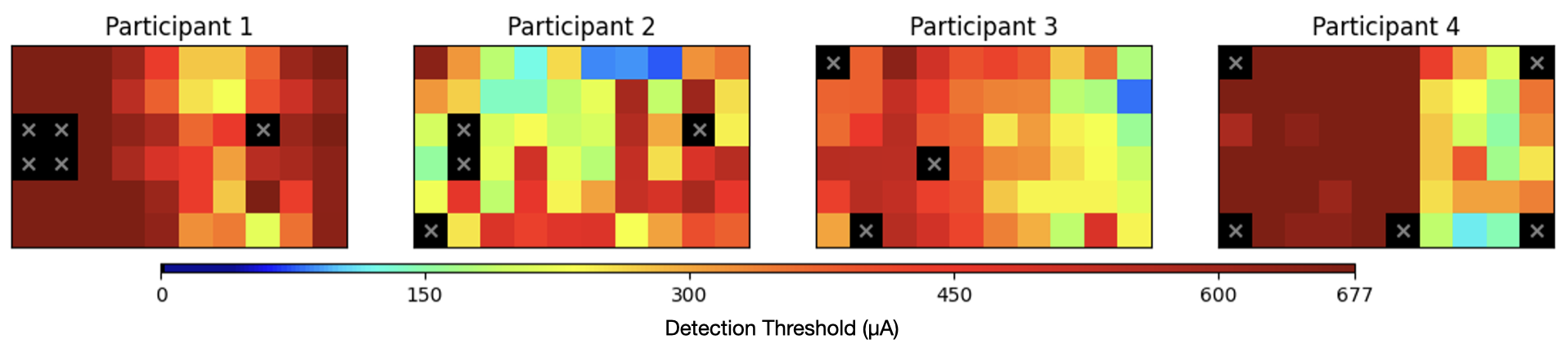}
    \caption{Perceptual threshold maps for four Argus II users, displayed on a $6 \times 10$ electrode array. Thresholds (in \SI{}{\micro\ampere}) were estimated using a Bayesian adaptive method~\cite{sadeghi_developing_2023} and are shown on a logarithmic color scale (dark blue = lowest, dark red = \SI{677}{\micro\ampere}). Inactive electrodes are marked with a cross.}
    \label{fig:thresholds}
\end{figure*}

To address this, we propose a framework based on \ac{GPR} to model the spatial structure of perceptual thresholds and reduce the number of required electrode measurements. 
While \ac{GPR} has been widely applied in spatial modeling across other domains~\cite{rasmussen_gaussian_2005}, this study represents the first application of \ac{GPR} to threshold estimation in visual prosthetics. By treating the threshold map as a continuous spatial field, GPR enables both threshold prediction at unsampled electrodes and uncertainty quantification, laying the foundation for data-efficient, adaptive sampling strategies.

As a proof of concept, we evaluate our framework using real-world threshold data from four Argus II users~\cite{sadeghi_developing_2023}. We systematically compare different GPR kernel functions to identify those best suited for modeling spatial variations in thresholds, and we benchmark several sampling strategies—including uniform random, spatially structured, and uncertainty-driven adaptive selection.

Our results demonstrate that GPR, combined with spatially optimized sampling, achieves highly accurate threshold predictions with far fewer measurements than traditional methods. Moreover, we find that while adaptive sampling can provide further refinements, spatial sampling alone often suffices, thereby offering a simple and robust solution for future high-density implants.

Beyond the context of retinal prostheses, this framework offers a generalizable strategy for adaptive calibration in other neuroprosthetic systems where stimulation thresholds exhibit spatial correlations.

\section{Methods}

\subsection{Dataset}

We used 240 previously published perceptual thresholds from four participants implanted with the Argus II epiretinal prosthesis~\cite{sadeghi_developing_2023} (Fig.~\ref{fig:thresholds}), which contains a $6 \times 10$ array of 60 electrodes~\cite{luo_argusr_2016}.
Perceptual thresholds were estimated using a Bayesian adaptive method in a yes/no procedure, as described in \cite{kontsevich_bayesian_1999}, where participants indicated whether they perceived a stimulus.

Each electrode required between 11 and 30 measurements to determine its threshold, ensuring reliable estimation while balancing participant fatigue. 
The current amplitude thresholds were estimated between \SI{40}{\micro\ampere} and \SI{677}{\micro\ampere} using 60 logarithmically spaced values. 
A maximum current of \SI{677}{\micro\ampere} was imposed due to device safety limits, meaning electrodes that required higher currents for perception were recorded as having the maximum value.

All other pulse parameters were fixed, with a pulse width of \SI{0.45}{\milli\second}, and a frequency of either \SI{6}{\hertz} or \SI{20}{\hertz} based on the participant's preference.

\subsection{Gaussian Process Regression}

To model the spatial variations in perceptual thresholds across the electrode array, we employed \acf{GPR}, a non-parametric Bayesian approach that provides a flexible framework for modeling spatially correlated data~\cite{rasmussen_gaussian_2005}.
In our framework, the covariance between any two points $\mathbf{x}_i$ and $\mathbf{x}_j$ is defined by a kernel function $k(\mathbf{x}_i,\mathbf{x}_j)$, which encodes assumptions about threshold variations across the implant.

Retinal prosthesis users exhibit complex spatial patterns in their perceptual thresholds (Fig.~\ref{fig:thresholds}), which can be attributed to factors such as electrode-retina distance, impedance variations, and localized retinal damage~\cite{ahuja_factors_2013,pogoncheff_explainable_2024}.
These patterns include both smooth global trends (e.g., due to gradual changes in tissue-electrode coupling) and sharp local discontinuities (e.g., due to scarring or nonfunctional electrodes).
To accommodate these diverse spatial structures, we evaluated three kernels $k(\mathbf{x}_i,\mathbf{x}_j)$, detailed below.

\subsubsection{Radial Basis Function (RBF) Kernel}
The RBF kernel assumes that perceptual thresholds vary smoothly across the array. It is defined as: \begin{equation} 
    k(\mathbf{x}_i, \mathbf{x}j) = \sigma^2 \exp\left( -\frac{d{ij}^2}{2\ell^2} \right), 
\end{equation}
where $\sigma^2$ is the signal variance (controlling the function’s overall variability),
$d_{ij}$ is the Euclidean distance between $\mathbf{x}_i$ and $\mathbf{x}_j$, and
$\ell$ is the length scale (which determines how quickly correlations decay with distance).
To balance flexibility and robustness, we initialized $\sigma^2$ in the range (0.1, 10) to prevent over-scaling and $\ell$ in the range (1, 50) to avoid overfitting to local noise.
In addition, a noise term $\sigma_n^2 \delta_{ij}$ was included in the kernel to account for experimental variability.

The RBF kernel is well-suited for modeling gradual changes in perceptual thresholds, such as those arising from electrode impedance variations or spatially consistent tissue-electrode coupling. 
However, its strong smoothness assumption makes it ill-equipped to handle sharp threshold discontinuities.

\subsubsection{Matérn Kernel}
The Matérn kernel extends GPR’s flexibility by allowing for non-smooth variations, making it better suited for threshold maps with sharp transitions. We used the Matérn kernel with $\nu = 1.5$, which is defined as: \begin{equation} k(\mathbf{x}i, \mathbf{x}j) = \sigma^2 \left( 1 + \frac{\sqrt{3} , d{ij}}{\ell} \right) \exp\left( -\frac{\sqrt{3} , d{ij}}{\ell} \right). \end{equation}

Like RBF, it includes $\sigma^2$ as the signal variance and
$\ell$ as the length scale, which controls how rapidly correlations decay.
We used the same parameter constraints as RBF, namely $\sigma^2 \in (0.1, 10)$, $\ell \in (1, 50)$, $\sigma_n^2 \delta_{ij}$ included for experimental noise.
The key difference is that the Matérn kernel does not enforce smoothness to the same degree as RBF, allowing it to model sharp threshold boundaries, such as those caused by localized scarring, poor electrode-retina contact, or damaged regions.

\subsubsection{Hybrid Matérn + RBF Kernel}

To leverage the advantages of both smooth and non-smooth models, we implemented a hybrid kernel:
\begin{equation}
    k(\mathbf{x}_i, \mathbf{x}_j) = \sigma^2 \left[ \alpha \, \text{RBF}(\ell_1) + (1-\alpha) \, \text{Matérn}(\ell_2) \right].
\end{equation}
Here, $\alpha$ is a weighting parameter that balances the RBF and Matérn contributions, and $\ell_1$ and $\ell_2$ represent separate length scales for smooth and discontinuous variations, respectively.

This hybrid approach defaults to RBF-like behavior ($\alpha \approx 1$) in areas where perceptual thresholds are smooth, but can shift toward Matérn-like behavior ($\alpha \approx 0$) where threshold discontinuities exist.
Therefore, the hybrid approach may be beneficial especially for implants with spatially mixed threshold structures, where some electrodes exhibit strong correlations while others show abrupt changes.

\begin{figure*}[!th]
    \centering
    \includegraphics[width=\linewidth]{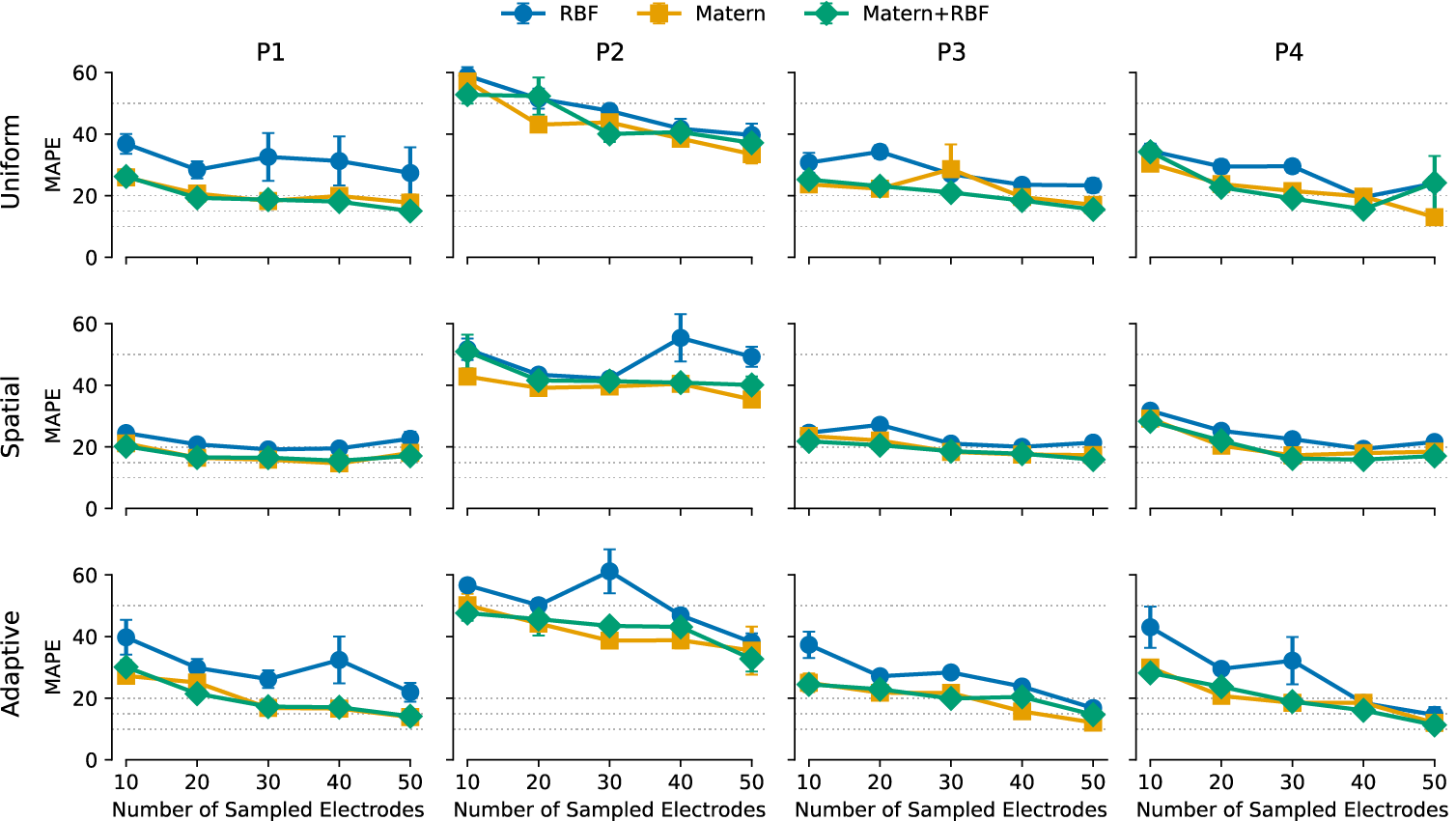}
    \caption{Mean absolute percentage error (MAPE) as a function of the number of sampled electrodes for four participants (P1–P4). Rows correspond to different sampling strategies (uniform, spatial, adaptive), while columns correspond to individual participants. Each plot compares the performance of three Gaussian Process Regression (GPR) kernels: Radial Basis Function (RBF, blue), Matérn (orange), and Matérn+RBF (green). Error bars indicate the standard error of the mean (SEM) across ten iterations. Dashed horizontal lines represent reference error levels (10, 15, 20, 50 MAPE).}
    \label{fig:results}
\end{figure*}

\subsubsection{Optimization and Evaluation}

Each \ac{GPR} model (with its open parameters $\sigma^2$ and $\ell$) was trained on a subset of available threshold data and tested on held-out electrodes to evaluate performance. Model hyperparameters were optimized by maximizing the log marginal likelihood (LML), a standard approach in GPR that ensures the learned function captures the underlying data structure effectively~\cite{rasmussen_gaussian_2005}.

To assess model accuracy, we computed the Mean Absolute Percent Error (MAPE), defined as:
\begin{equation}
    \text{MAPE} = \frac{1}{N} \sum_{i=1}^{N} \left| \frac{y_i - \hat{y}_i}{y_i} \right| \times 100.
\end{equation}
This metric quantifies the relative deviation of predicted thresholds ($\hat{y}_i$) from actual measured values ($y_i$), making it robust to variations in absolute threshold levels. Unlike standard mean absolute error, MAPE accounts for the relative importance of different electrodes, particularly in cases where thresholds span multiple orders of magnitude.

Importantly, \ac{GPR} provides both point estimates and uncertainty measures for each prediction. The predictive variance serves as a confidence metric, which can be used to guide adaptive sampling strategies by prioritizing high-uncertainty regions for measurement.


\subsection{Electrode Sampling Strategies}

To efficiently estimate perceptual thresholds while minimizing the number of sampled electrodes, we employed three distinct sampling strategies: uniform random sampling, spatial sampling, and adaptive variance-based sampling. The choice of sampling strategy directly impacts the model’s ability to reconstruct the full threshold map while reducing measurement burden.

\subsubsection{Uniform Random Sampling}
Uniform random sampling selects electrodes randomly from the valid set without replacement. This ensures an unbiased selection process but can lead to uneven spatial coverage, with electrodes clustering in certain regions while leaving others undersampled. Such clustering can degrade model performance by failing to capture large-scale spatial trends in threshold variations.
Despite its limitations, uniform sampling serves as a baseline against which more structured sampling approaches can be compared.

\subsubsection{Spatial Sampling}
To mitigate the uneven coverage issues of random selection, we implemented a spatial sampling strategy based on \ac{PDS}. This method enforces a minimum spacing constraint, ensuring that sampled electrodes are evenly distributed across the array.

\Ac{PDS} begins with a randomly selected seed electrode. Subsequent electrodes are iteratively chosen to maximize pairwise distances, preventing excessive clustering. Candidates are proposed within an annular region around previously sampled electrodes, and each is accepted only if it remains sufficiently distant from all previously chosen points. If no valid candidates remain after a predefined number of attempts, the sampling process terminates.

By ensuring even spatial coverage, this method improves the ability of the model to interpolate across unsampled locations, particularly when the number of sampled electrodes is small.

\subsubsection{Adaptive Variance-Based Sampling}
In addition to structured spatial sampling, we implemented an adaptive sampling procedure that dynamically selects electrodes based on model uncertainty, leveraging \ac{GPR}'s predictive variance.

This procedure begins by training an initial GPR model on a small, randomly selected subset of electrodes. Using this model, we estimate perceptual thresholds for the remaining electrodes and compute the predictive uncertainty (i.e., the standard deviation of the posterior distribution at each electrode). Electrodes with the highest uncertainty are then prioritized for subsequent measurements, as sampling these locations is expected to provide the most informative data for refining the threshold map.
This iterative process continues until a predefined number of electrodes have been sampled. At each step, the model is updated incrementally with new measurements, reducing uncertainty and improving threshold estimates across the array.

Unlike uniform or spatial sampling, this variance-driven approach is adaptive, meaning the locations selected depend on the evolving structure of the threshold map rather than a fixed spatial pattern.
This procedure may be most beneficial when threshold variations are highly localized, but it may not always outperform spatial sampling in cases where thresholds are smoothly distributed.

\subsection{Statistical Analysis}

For statistical comparisons of kernel and sampling strategies, we used the Wilcoxon signed-rank test, a non-parametric method suitable for paired, non-normally distributed data. P-values were computed separately for each participant and across all participants combined, with a significance threshold of $p<.05$.

\section{Results}

\subsection{Effect of Kernel Choice on Prediction Accuracy}

Fig.~\ref{fig:results} shows the performance of different \ac{GPR} models across four participants, comparing the Matérn, RBF, and Matérn+RBF kernels as a function of the number of sampled electrodes. Each model was trained using either uniform, spatial, or adaptive sampling, with mean absolute percentage error (MAPE) as the primary evaluation metric.  

Across all participants, the Matérn kernel consistently outperformed RBF, yielding lower MAPE values across most sample sizes ($p<.001$, Wilcoxon signed-rank test). 
The RBF kernel exhibited greater variance and fluctuations in accuracy, suggesting increased sensitivity to sample selection and overfitting when the training set was unevenly distributed. 

At the participant level, the Matérn kernel significantly outperformed RBF in all four cases (Participant 1: $p<.001$, Participant 2: $p<.01$,  Participant 3: $p<.05$, Participant 4: $p<.001$).
The Matérn+RBF hybrid did not improve upon the standard Matérn kernel in any participant ($p>.05$).

Beyond kernel comparisons, MAPE trends across sample sizes highlight the feasibility of reducing electrode measurements without major performance loss. When sampling as few as 20 electrodes, models trained with a Matérn kernel achieved MAPE values below 20\%, with further improvements as more electrodes were sampled. In optimal conditions, where spatial sampling was used with the Matérn kernel, MAPE dropped below 10\% at 50 sampled electrodes, suggesting that accurate threshold estimation can be achieved with fewer measurements than traditional exhaustive testing.

These results establish the Matérn kernel as the most effective choice for modeling perceptual thresholds, offering both stability and flexibility. The failure of the hybrid Matérn+RBF kernel to provide additional benefits suggests that the standard Matérn formulation is sufficient for capturing the spatial structure of perceptual thresholds without unnecessary parameter tuning.

\subsection{Effect of Sampling Strategy on Prediction Accuracy}

Across all participants, spatial sampling significantly outperformed uniform sampling ($p<.01$), with the greatest improvements observed when fewer electrodes were sampled.
At the participant level, spatial sampling was significantly better than uniform in Participants 1 ($p<.001$) and 3 ($p<.05$), while no significant differences were observed in Participants 2 and 4.

Adaptive sampling, designed to iteratively select high-uncertainty electrodes, performed similarly to uniform sampling across all participants ($p=.074$), except for Participant 1, where it showed a marginal advantage over spatial sampling ($p<.05$).

The relationship between sampling strategy and MAPE follows a similar pattern. Uniform sampling generally resulted in higher MAPE, particularly when fewer than 20 electrodes were sampled. Spatial sampling provided the most consistent improvements, with MAPE dropping below 20\% at 20 electrodes and continuing to decline as more electrodes were sampled. Adaptive sampling, while promising in theory, did not consistently improve over spatial sampling in this dataset.

These results indicate that structured spatial sampling is a robust and effective method for threshold estimation, consistently improving accuracy over random selection and performing at least as well as adaptive sampling. While adaptive methods hold promise for larger electrode arrays where uncertainty-based selection could play a greater role, they did not provide a measurable advantage in this dataset.

\section{Discussion}

In this study, we introduced a \acf{GPR} framework~\cite{rasmussen_gaussian_2005} to efficiently estimate perceptual thresholds across retinal implant electrode arrays, reducing the number of required measurements while maintaining high predictive accuracy. Our findings demonstrate that spatially optimized sampling consistently outperformed uniform random sampling, providing a structured method for reducing patient burden without sacrificing accuracy. While adaptive sampling, which prioritizes high-uncertainty regions, showed promise, its improvements over spatial sampling were not statistically significant in this dataset.

\Ac{GPR} effectively captured the spatial correlations in perceptual thresholds, allowing for accurate predictions at unsampled locations. The Matérn kernel consistently outperformed the RBF kernel, indicating that a model capable of handling sharp threshold discontinuities is necessary for accurate interpolation across the electrode array. 
This finding is consistent with the known variability in perceptual thresholds~\cite{de_balthasar_factors_2008, shivdasani_factors_2014}, which can exhibit both smooth and abrupt spatial variations due to factors such as electrode-neuron distance, local retinal health, axonal stimulation, and tissue-electrode interactions~\cite{yucel_factors_2022,hou_axonal_2024}.
In addition, non-local activation along ganglion axon pathways can lead to spatial distortions in percepts~\cite{beyeler_model_2019}, further motivating the need for flexible models like GPR that can accommodate both local and extended spatial correlations.

In terms of sampling strategy, spatially optimized sampling proved to be the most robust approach, particularly when the number of sampled electrodes was small. By ensuring an even distribution of selected electrodes, this method minimized the risk of overfitting to localized clusters and provided stable error convergence. Adaptive sampling, while conceptually appealing, did not significantly improve prediction accuracy over spatial sampling. This suggests that, in a structured implant array with a moderate number of electrodes (e.g., 60 in Argus II~\cite{luo_argusr_2016}), spatial sampling already provides an efficient and effective way to estimate thresholds. However, adaptive sampling may still hold value in larger electrode arrays where simple spatial spacing may not be sufficient to capture localized threshold variations.

Despite these promising findings, some challenges remain. First, while adaptive sampling did not outperform spatial sampling in our dataset, it is possible that larger implants or more heterogeneous threshold maps could reveal stronger benefits. In such cases, a hybrid strategy (i.e., starting with spatial sampling and refining with adaptive sampling) could provide the best of both worlds. Second, GPR is ultimately limited by the reliability of perceptual reports, which can be influenced by fatigue, attention, and intra-session variability~\cite{yue_ten-year_2015,d_signal_2012,lu_characterizing_1999}.

While this study focused on spatial correlations in perceptual thresholds, future extensions could incorporate additional physiological or anatomical factors to further enhance model accuracy. Prior work has identified electrode impedance, retinal eccentricity, and demographic factors (e.g., subject age, time since blindness onset) as important predictors of threshold variability~\cite{pogoncheff_explainable_2024}. Such features could be integrated into the GPR framework by expanding the input space to include electrode-specific impedance measurements, anatomical distances, or patient history, allowing the model to learn both spatial and non-spatial correlations. Alternatively, these factors could inform kernel design or serve as priors over expected threshold distributions. Although integrating multimodal data was beyond the scope of the present study, doing so offers a principled, interpretable path toward more personalized and scalable calibration strategies.

In conclusion, this study establishes GPR with a Matérn kernel and spatial sampling as an effective approach for streamlining the calibration process of retinal implants. While adaptive sampling remains an intriguing avenue for future work, spatial sampling alone was sufficient for efficient threshold estimation in the current dataset. As next-generation prosthetic implants move toward higher electrode densities~\cite{palanker_photovoltaic_2020, chenais_photovoltaic_2021,musk_integrated_2019}, intelligent sampling strategies like these will be essential for reducing patient burden and making clinical calibration procedures more feasible.

\bibliographystyle{IEEEtran}
\bibliography{references}

\end{document}